\documentclass[epj]{webofc}
\usepackage[varg]{txfonts}   
\usepackage{amsfonts}
\usepackage{booktabs}
\usepackage{lineno}
\usepackage{multirow}
\usepackage{amsmath} 
\usepackage{graphicx}
\usepackage{subfigure}
\begin{document}
%
%
\title{Towards a reconstruction of Supernova Neutrino Spectra in JUNO}
%
%

\author{\textit{Cristina} Martellini\inst{1,2}\fnsep\thanks{\email{cristina.martellini@uniroma3.it}}, 
        \textit{Stefano} Maria Mari\inst{1,2}, 
        \textit{Paolo} Montini\inst{1,2} \and 
	\textit{Giulio} Settanta\inst{1,2}
}

\institute{Dipartimento di Matematica e Fisica, Universita' degli Studi di Roma Tre, Via della Vasca Navale 84, 00146 Rome, Italy 
\and
           INFN Sezione di Roma Tre, Via della Vasca Navale 84, 00146 Rome, Italy 
          }
\abstract{%
  Observation of supernovae (SN) through their neutrino emission is a fundamental point to understand both SN dynamics and neutrino physical properties. JUNO is a 20kton liquid scintillator detector, under construction in Jiangmen, China. The main aim of the experiment is to determine neutrino mass hierarchy by precisely measuring the energy spectrum of reactor electron antineutrinos. However due to its properties, JUNO has the capability of detecting a high statistics of SN events too. Existing data from SN neutrino consists only of 24 events coming from the SN 1987A,the detection of a SN burst in JUNO at $\sim 10 kpc$ will yield $\sim 5 x 10^{3}$ inverse beta decay (IBD) events from electron antineutrinos, about 1500 from proton elastic scattering (pES) above the threshold of 0.2 MeV, about 400 from electron elastic scattering (eES), plus several hundreds on other CC and NC interaction channels from all neutrino species.}
\maketitle
\section{Introduction}
\label{intro}
The Jiangmen Underground Neutrino Observatory (JUNO)\cite{An:2015jdp} is a 20kton multi-purpose underground liquid scintillator (LS) detector, which is located at Kaiping, Jiangmen in South China, is design primarily to determine the neutrino mass hierarchy (MH) using reactor antineutrinos.

The detector features make it suitable for a whole series of measurements in the $\nu$ Physics such as atmospheric neutrinos, geoneutrinos, solar neutrinos and supernova and diffuse supernova neutrinos, as a natural accessible source and other exotic searches\cite{An:2015jdp}.The JUNO central detector consists of a $\sim 36 m$ diameter acrylic sphere, filled with 20kton of liquid scintillator (LS). The light will be collected by a double-system of photosensors: 18.000 20" PMTs and 25.000 3" PMTs. As a LS calorimeter it has a very low energy threshold and can measure supernova neutrinos with unprecedent performaces and excellent energy resolution. In this work, we present simulations studies on Core Collapse Supernova (CCSN) neutrino event for different detection channels involving different flavours of SN neutrinos. Using the supernova flux model from Nakazato \textit{et al} \cite{Nakazato:2015rya} and assuming the galactic supernova explosion occurring at three different distances, a statistic sample can be simulated to be able to build an unfolding method to get back to the original SN parameters and hence reconstruct the SN neutrino energy spectra. 
\section{Core Collapse Supernova Neutrinos}
\label{sec-1}
A massive star of mass above $\sim 8 M_{\odot}$ is expected to experience a core collapse under its own gravity and then a violent explosion, where 99$\%$ of the gravitational binding energy of the core-collapse supernova (CCSN) will be carried by the intense burst of neutrinos. Because of this they play a decisive role during all the stages of such an event. While the neutrinos detected from the SN 1987A were useful to give basic confirmation of the theory of neutron star (NS) formation, their event statistics was too poor to highlight some useful information about the dynamics of the explosion mechanism of the CCSN\cite{Mirizzi:2015eza}. High-statistics measurement of neutrinos from the next Galactic SN will therefore be of major importance for astrophysics, but also neutrino and nuclear physics. As the largest new generation LS detector, JUNO will be superior in its high statistics, in the energy resolution (3$\%/\sqrt{MeV}$) and flavour informations. Since the average SN distance is around 10 kpc, JUNO will register about 5000 from inverse beta decay (IBD) caused by the interaction of electron antineutrinos with the LS, (${\bar{\nu}}_{e}+p \rightarrow n + e^{+}$), around 1500 events from all-flavour elastic neutrino-proton elastic scattering, ($\nu +p \rightarrow \nu + p$), more than 300 events from neutrino-electron elastic scattering, ($\nu + e^{-} \rightarrow \nu + e^{-}$), as well as other charged current (CC) and neutral current (NC) interactions on the $^{12}C$ nuclei.

\begin{table}[htpb]
\centering
\caption{Numbers of neutrino events in JUNO for a SN at a typical distance of 10 kpc. Three representative values of the average neutrino energy $\langle E_{\nu} \rangle$ = 12, 14 and 16 MeV are taken for illustration. For the elastic neutrino-proton scattering a threshold of 0.2 MeV for the proton recoil energy is considered.}
\label{tab-1}       
\begin{tabular}{lcccc}
\hline
Channel & Type & \multicolumn{3}{c}{ Events for different $\langle E_{\nu} \rangle$ values} \\
\cmidrule(lr){3-5}
  &    & 12 MeV &  14 MeV & 16 MeV \\
\midrule
${\bar{\nu}}_{e}+p \rightarrow e^{+}+n $ & CC & $4.3\times10^{3}$ & $5.0\times10^{3}$ & $5.7\times10^{3}$ \\
$\nu + p \rightarrow \nu + p$ & NC  & $0.6 \times10^{3}$ & $1.2\times10^{3}$ & $2.0\times10^{3}$ \\
$\nu + e \rightarrow \nu + e$ & ES  & $3.6\times10^{2}$ & $3.6\times10^{2}$  & $3.6\times10^{2}$ \\
$\nu + ^{12}C \rightarrow \nu + ^{12}C^{*}$ & NC  & $1.7\times10^{2}$ & $3.2\times10^{2}$  & $5.2\times10^{2}$ \\
$\nu_{e} + ^{12}C \rightarrow e^{-} + ^{12}N$ & CC  & $0.5\times10^{2}$ & $0.9\times10^{2}$  & $1.6\times10^{2}$ \\
${\bar{\nu}}_{e} + ^{12}C \rightarrow e^{+} + ^{12}B$ & CC  & $0.6\times10^{2}$ & $1.1\times10^{2}$  & $1.6\times10^{2}$ \\\hline
\end{tabular}
\vspace*{0.5cm}  
\end{table}

The numbers of neutrino events at JUNO\cite{An:2015jdp} for a SN at a typical distance of 10 kpc are presented in Table \ref{tab-1}. 
With these measurements of SN neutrinos, JUNO may have the capability to get unique informations to measure the initial SN neutrino fluxes, to constrain the neutrino's mass scale and ordering, to test the scenario of collective neutrino oscillations and even to probe the neutrino electromagnetic properties. 

\section{Monte Carlo Supernova Simulations}
\label{sec-2}
\begin{figure}[htbp]
\centering%
\subfigure[\protect\url{Visible \ Energy}\label{fig:airplane}]%
{\includegraphics[width=60mm]{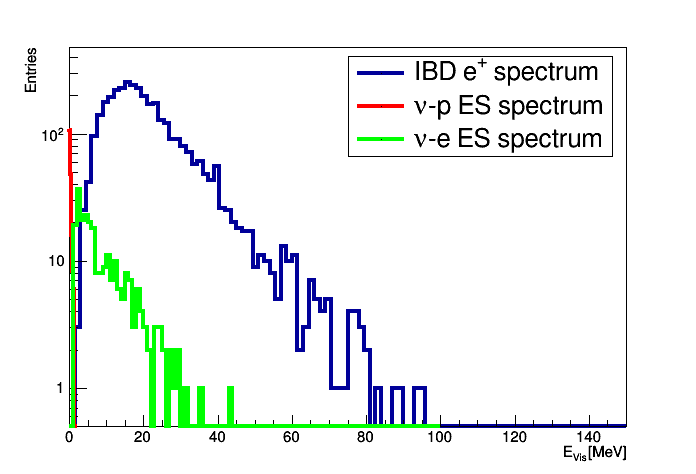}}\qquad\qquad
\subfigure[\protect\url{PE distribution}\label{fig:lena}]%
{\includegraphics[width=60mm]{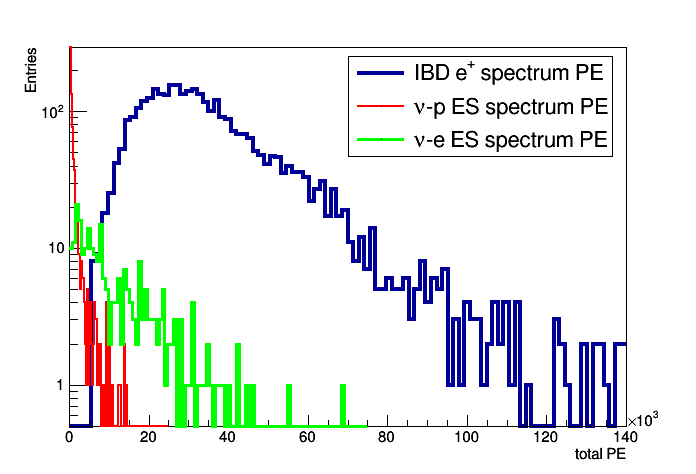}}\qquad\qquad
\caption{Left: Distribution of the energy spectra as a function of the visible energy for IBD (blue),$\nu-p$ ES (red) and $\nu-e$ ES (green). Right: distribution of the NPE for the IBD (blue), $\nu-p$ ES (red) and $\nu-e$ ES (green). \label{fig:fig-1}}
\end{figure}

For the simulation the Supernova Generator implemented in the JUNO Software has been used, setting a sample based on different solar masses ($M_{\odot}$), metallicity (Z) and revival time ($\tau_{rev}$), setting our simulation supposing normal hierarchy (NH), we generated SN at three different reference distances, building the statistic sample shown in Table\ref{tab-1}. 
\begin{table}
\centering
\caption{Different SN fluxes generated by different progenitor star at three different distances and with two values of the metallicity.}
\label{tab-2}       
\begin{tabular}{cccc}
\hline
Mass    & Metallicity (Z) & Revival time $\tau$ (ms) & Distance D (kpc)  \\\hline
\multirow{3}{*}{13 $M_{\odot}$} & \multirow{2}{*}{0.02} & \multirow{3}{*}{300} & 2 \\
 & \multirow{2}{*}{0.04} &  & 10 \\
 &  &  &  20 \\\hline
\multirow{3}{*}{20 $M_{\odot}$} & \multirow{2}{*}{0.02} & \multirow{3}{*}{300} & 2\\
 & \multirow{2}{*}{0.04} & & 10\\
 &  &  & 20 \\\hline
\multirow{3}{*}{50 $M_{\odot}$} & \multirow{2}{*}{0.02} & \multirow{3}{*}{300} & 2 \\
 & \multirow{2}{*}{0.04} &  & 10 \\
 &  &  & 20 \\\hline
\end{tabular}
\vspace*{0.5cm}  
\end{table}
We simulated therefore 18 SN bursts, which provided us with different event rates and then we run those through the detector to evaluate its response. We plotted the energy spectrum from each of the different channels first as a function of the visible energy and subsequently as a function of the number of photoelectrons (PE) detected by the two PTMs optical systems in JUNO. In Figure \ref{fig:fig-1} below the distributions for all the different channels are shown.

\section{Unfolding Method}
\label{sec-3}
\vspace{1mm}
Following the analysis of the simulation the main aim of this work is to build an unfolding method to get back to the original parameters of the SN and reconstruct the original neutrino spectra. Taking into consideration the  contributions given to their event rate in the detector, the three main channels have been taken into account for the development of this work, which are the IBD event, the $\nu-p$ ES and the $\nu-e$ ES event. Depending on the channel, the probability of having a $\nu-{\bar{\nu}}$ of any flavour of a given energy have been identified as follow:
\begin{equation}
\begin{aligned}
P_{IBD}(E_{{\bar{\nu}}_{e}}) &\propto \int\limits_{E_{min}}^{E_{max}} P_{IBD}(E_{{\bar{\nu}}_{e}}|E_{0})\cdot P_{IBD}(E_{0}) \mathrm{d}E_{0} \\ 
P_{pES}(E_{\nu}) &\propto \sum_{flavour=1}^{3} \int\limits_{E_{min}}^{E_{max}} P_{pES}^{flavour}(E_{\nu}|E_{0})\cdot P_{pES}^{flavour}(E_{0}) \mathrm{d}E_{0} \\ 
P_{eES}(E_{\nu}) &\propto \sum_{flavour=1}^{3} \int\limits_{E_{min}}^{E_{max}} P_{eES}^{falvour}(E_{\nu}|E_{0})\cdot P_{eES}^{flavour}(E_{0}) \mathrm{d}E_{0} \\
\end{aligned}
\end{equation}  
where we consider the probability of measuring the energy $E_{0}$ in the detector $P_{IBD}(E_{0})$, $P_{pES}^{flavour}(E_{0})$ and $P_{eES}^{flavour}(E_{0})$ as the observables, the $P_{IBD}(E_{{\bar{\nu}}_{e}})$, $P_{pES}^{flavour}(E_{\nu})$ and $P_{pES}^{flavour}(E_{\nu})$ as the unfolded spectra which are given by the convolution of the prior probabilities and the conditional probabilities that given the visible energy $E_{0}$ measured by the JUNO detector, this has been generated by $\nu/ \bar{\nu}$ of energy $E_{\nu}$. These last ones are obtained from the Bayes Theorem. 
The expressions are then integrated within an energy range ($E_{min}, E_{max}$) and for the two elastic scattering also sum on the three different neutrino flavours. 
\section{Conclusions}
\label{sec-5}
This work illustrates a preliminary study on the capability of the JUNO experiment to detect SN neutrinos and to act together with other neutrino experiment as an alert system for astrophysical object. How reconstructing the SN energy neutrino spectra gives us the chance to learn useful informations about the SN evolution mechanism has been appointed. JUNO should be competitive and improve our knowledge not just on the neutrino oscillation parameters but also with a unique potential for SN neutrinos.


%
\bibliography{biblio.bib}
%

\end{document}